\documentclass[12pt]{iopart}
\usepackage{epsfig}
\usepackage{amssymb}

\begin{document}

\title{Instabilities, solitons, and rogue waves in $\mathcal{PT}$-coupled
nonlinear waveguides}
\author{ Yu V Bludov$^{1}$, R Driben$^{2,3}$, V V Konotop$^{4}$ and B A
Malomed$^{2}$}

\address{$^{1}$Centro de F\'{i}sica, Universidade do Minho,
Campus de Gualtar, Braga 4710-057, Portugal}

\address{$^{2}$Department of Physical Electronics, School of Electrical Engineering,
Faculty of Engineering, Tel Aviv University, Tel Aviv 69978, Israel}

\address{$^{3}$Department of Physics, University of Paderborn, Warburger
Str. 100, D-33098 Paderborn, Germany}

\address{$^{4}$Centro de F\'{i}sica Te\'{o}rica
e Computacional and Departamento de F\'{i}sica, Faculdade de Ci\^encias, Universidade
de Lisboa, Avenida Professor Gama Pinto 2, Lisboa 1649-003, Portugal}

\ead{bludov@fisica.uminho.pt} \ead{konotop@cii.fc.ul.pt}

\begin{abstract}
We considered the modulational instability of continuous-wave  backgrounds,
and the related generation and evolution of deterministic rogue waves in the
recently introduced parity-time ($\mathcal{PT}$)-symmetric system of
linearly-coupled nonlinear Schr\"{o}dinger equations, which describes a
Kerr-nonlinear optical coupler with mutually balanced gain and loss in its
cores.  {Besides the linear coupling, the overlapping cores
are coupled through cross-phase-modulation term too.} While the rogue waves,
built according to the pattern of the Peregrine soliton, are (quite
naturally) unstable, we demonstrate that the focusing cross-phase-modulation
interaction results in their partial stabilization. For $\mathcal{PT}$%
-symmetric and antisymmetric bright solitons, the stability region is found
too, in an exact analytical form, and verified by means of direct
simulations.
\end{abstract}

\maketitle



\section{Introduction}

It is a generally recognized fact that, independently of the underlying
physics, an instability of the background is a prerequisite for the
emergence of regular or random rogue waves (see, e.g., the discussion
in~Ref. \cite{special}). In its turn, the instability is determined, on the
one hand, by the interplay between the dispersion and nonlinearity, and, on
the other hand, by the competition between losses and gain, if an open
system is considered. In this latter case, one can speak about \textit{%
dissipative rogue waves}~\cite{dissipative}, which are identified by an
enhanced probability of generating high-amplitude pulses.

In addition to the above-mentioned generic situations, there exist special
dissipative systems obeying the so-called parity-time ($\mathcal{PT}$)
symmetry, i.e., featuring spatially separated and exactly balanced gain and
loss. These systems are described by non-Hermitian Hamiltonians, which may
have purely real spectra of eigenvalues, provided that the strength of the
anti-Hermitian part of the Hamiltonian (which accounts for the balanced gain
and loss) does not exceed a certain critical value~\cite%
{bender1998,c:bender2007}.

 {Optics} represents a unifying framework for a variety of
wave phenomena. In particular, the $\mathcal{PT}$-symmetry was
experimentally implemented in  {coupled optical
waveguides~\cite{c:pt-experim}. Moreover principles of its implementation in
plasmonic waveguides~\cite{plasmonic} and in a gaseous mixtures of resonant
atoms~\cite{atomic}, were recently proposed.} On the other hand, optical
rogue waves have also been observed in some settings~\cite{rwopt,AkhNatPhys}
and predicted in others, such as periodic arrays of waveguides~\cite{BKA_OL}%
. While the original ideas of the use of the $\mathcal{PT}$ symmetry in
quantum mechanics imply complex potentials obeying condition $V(x)=\overline{%
V(-x)}$ \cite{c:bender2007} (hereafter the overbar stands for complex
conjugation), in the experimental realization~\cite{c:pt-experim} and
numerous theoretical studies nonlinear dual-core waveguides (\textit{couplers%
}), with one core carrying the gain and the other one being lossy, were
explored as an optical implementation of the $\mathcal{PT}$-symmetric
systems. The dual-core systems are described by systems of coupled nonlinear
Schr\"{o}dinger equations (NLSEs), one with the gain and the other --- with
loss. These models and their generalizations in a form of sequence of
couplers give rise to bright~\cite{c:solitons,c:PTepl,barashenkov1,necklace}
and dark~\cite{BKM_dark} solitons, vortices~\cite{Leykam}, breathers~\cite%
{barashenkov2}, and describe a switch for solitons between the cores~\cite%
{AKOS}.

As concerns optical rogue waves, there are two major directions of the work
in this field. The first relates rogue events to the well-known process of
supercontinuum generation \cite{Ranka, Herrmann, Dudleyobzor, Skryabinobzor,
SCnano} in optical fibers. The soliton dynamics affects the supercontinuum
generation process at a very early stage, \textit{viz}., the fission \ of
higher-order solitons \cite{Satsuma, Herrmann, DMSY}, which is followed by
multiple interactions of solitons with dispersive waves at advanced stages
\cite{Skryabinobzor, Yulin, DribenMitschke}. In particular, the
strongest-Raman-shifted solitons \cite{Mitschke} were proposed as possible
candidates for rogue waves \cite{rwopt}. Crests of soliton collisions were
proposed too, as alternative candidates \cite{DudleyGenty, Genty}. Recently,
``long-lasting" accelerating optical rogue waves with an oblong shape,
resembling the shape of their oceanic counterparts, were reported \cite%
{DribenBabushkin, Demircan}. Another approach \cite{Dudley1, Akhmet, Akhmet2}
is based on solutions for Akhmediev breathers \cite{AB}, and, in particular,
on the single-peak solution  {often referred to as the
Peregrine soliton (or Peregrine rogue wave)} \cite{c:peregrine}, which
represents a \textit{deterministic rogue wave}~\cite{c:akhmediev2009}
generated by the NLSE recently observed experimentally~\cite{AkhNatPhys}.
These works reveal waves which arise from the modulational instability (MI)
and subsequently disappear, which is consistent with the behavior of the
famous ship killers in the ocean \cite{killer}.

The main objective of the present paper is to study rogue waves in $\mathcal{%
PT}$-symmetric optical models based on the dual-core couplers. One of our
goals is to introduce an analog of the Peregrine soliton in this setting.
More specifically, we are interested in how the presence of the balanced
dissipation and gain, i.e., the $\mathcal{PT}$ symmetry, affects the MI of
the background and possibility of the creation of waves localized in space
and time in such systems. In this context, it is relevant to mention a
number of previous studies of the deterministic rogue waves carried out in
the framework of the coupled NLSEs describing two-component matter waves in
Bose-Einstein condensates~\cite{c:rw-BEC2}, multi-parametric vector
solitons, and, in particular, bright-dark-rogue waves~\cite{degasperis}.

The rest of the paper is organized as follows. The model is introduced in
Section II, which is followed by the analysis of the MI of the
continuous-wave (CW) solutions in Section III, and the study of rogue-wave
solutions, following the pattern of the Peregrine soliton, in Section IV.
Exact analytical results, verified by direct simulations, for the stability
of $\mathcal{PT}$-symmetric and antisymmetric solitons in the same system
are reported in Section V, and the paper is concluded by Section VI.

\section{The model}

We consider a system of linearly coupled NLSEs for field variables $\psi _{1}
$ and $\psi _{2}$:
\begin{eqnarray}
i\frac{\partial \psi _{1}}{\partial z} &=&-\frac{\partial ^{2}\psi _{1}}{%
\partial x^{2}}+\left( \chi _{1}|\psi _{1}|^{2}+\chi |\psi _{2}|^{2}\right)
\psi _{1}+i\gamma \psi _{1}-\psi _{2},  \label{eq:NLS_gen1} \\
i\frac{\partial \psi _{2}}{\partial z} &=&-\frac{\partial ^{2}\psi _{2}}{%
\partial x^{2}}+\left( \chi |\psi _{1}|^{2}+\chi _{1}|\psi _{2}|^{2}\right)
\psi _{2}-i\gamma \psi _{2}-\psi _{1}.  \label{eq:NLS_gen2}
\end{eqnarray}%
which describes a set of two parallel planar waveguides, with $z$ and $x$
being dimensionless propagation and transverse coordinates. Accordingly, the
initial-value problem corresponds to an optical beam shone into the
waveguides input at given $z=z_{i}$. Alternatively, the model describes a
dual-core fiber coupler, where $x$ plays the role of the temporal variable
\cite{c:solitons,c:PTepl,AKOS,barashenkov1}. Equations (\ref{eq:NLS_gen1})
and (\ref{eq:NLS_gen2}) are coupled nonlinearly by the cross-phase
modulation (XPM) $\sim \chi $, and linearly by the last terms with
respective coupling constant scaled to be $1$. Lastly, constant $\gamma >0$
describes the $\mathcal{PT}$-balanced gain in Eq. (\ref{eq:NLS_gen1}) and
dissipation in Eq. (\ref{eq:NLS_gen2}). In optics, this setting can be
realized using a system of two lossy parallel-coupled waveguides, doped by
gain-providing atoms, in which only one waveguide is pumped by the external
source of light providing the gain.

Although the first core carries the gain, its linear coupling to the lossy
mate makes the zero state in the system neutrally stable, allowing for
propagation of linear waves. This is the well-known situation, which takes
place if the gain/loss term is small enough, compared to the linear coupling
through which the energy is transferred from the core with the gain to the
lossy one, or, more specifically, when $\gamma \leq 1$ \cite{Javid}. In such
a situation, modes can be excited in the system by input beams but do not
arise spontaneously. Below, without the loss of generality, we restrict the
consideration to this case, and therefore introduce a convenient
parametrization,
\begin{equation}
\gamma =\sin \delta ,~~0<\delta <\pi /2.  \label{delta}
\end{equation}

Following~Ref. \cite{c:solitons}, we look for $\mathcal{PT}$-symmetric ($+$)
and antisymmetric ($-$) solutions to Eqs. (\ref{eq:NLS_gen1}) and (\ref%
{eq:NLS_gen2}) as%
\begin{equation}
\psi _{2}\left( x,z\right) =\pm e^{\pm i\delta }\psi _{1}\left( x,z\right) ,
\label{21}
\end{equation}%
with function $\psi _{1}$ obeying the single equation,%
\begin{equation}
i\frac{\partial \psi _{1}}{\partial z}=-\frac{\partial ^{2}\psi _{1}}{%
\partial x^{2}}+\left( \chi _{1}+\chi \right) |\psi _{1}|^{2}\psi _{1}\mp
\left( \cos \delta \right) \psi _{1}.  \label{1}
\end{equation}%
An observation particularly relevant to the solutions having the form of Eq.
(\ref{21}) is that the dissipation and gain break the conventional symmetry
of the coupler {. The conventional symmetry is now substituted
by the following reduction:} if $\left( \psi _{1}(x,z),\psi _{2}(x,z)\right)
$ is a solution of Eqs. (\ref{eq:NLS_gen1}) and (\ref{eq:NLS_gen2}), then
pair $\left( \overline{\psi }_{2}(x,-z),\overline{\psi }_{1}(x,-z)\right) $
is a solution too. This reduction corresponds to the change $\delta
\rightarrow \pi -\delta $. Therefore, below we consider the domain of the
variation of $\delta $ to be $[0,\pi ]$, where values $\delta $ and $\pi
-\delta $ corresponds to the two different solutions at the same dissipation
and gain. In other words, intervals $0\leq \delta \leq \pi /2$ and $\pi
/2\leq \delta \leq \pi $ correspond to the $\mathcal{PT}$-symmetric and $%
\mathcal{PT}$-antisymmetric solutions.

\section{Modulational instability}

Up to a trivial phase shift, CW solutions to Eqs. (\ref{eq:NLS_gen1}) and
Eq. (\ref{eq:NLS_gen2}) are
\begin{equation}
\psi _{j}^{(\mathrm{cw})}=\rho \exp \left[ ikx-ibz+i(-1)^{j}\delta /2\right]
,  \label{eq:pl-wave-cur}
\end{equation}%
where $k$ represents a background current, and $b=k^{2}+\rho ^{2}(\chi
_{1}+\chi )-\cos \delta $ (see, e.g., \cite{BKM_dark} for more details),
i.e., the amplitudes of the fields are equal in both cores, which is natural
in view of the necessity to ensure the balance between gain and loss. To
study the MI of the CW states, we use the standard ansatz,
\[
\psi _{j}=\rho \left[ e^{i(-1)^{j}\delta /2}+\eta _{j}e^{-i(\beta z-\kappa
x)}+\bar{\nu}_{j}e^{i(\bar{\beta}z-\kappa x)}\right] e^{ikx-ibz},
\]%
$j=1,2$, with $|\eta _{j}|\,|\nu _{j}|\ll  {1} $. Then, two
branches $\beta =\beta _{1,2}(k)$ of the dispersion relation for the
stability eigenvalues are given by 
\begin{eqnarray}
&&\beta _{1}(\kappa )\equiv 2k\kappa \pm \kappa \sqrt{\kappa ^{2}+2\rho
^{2}(\chi _{1}+\chi )},  \label{omega1} \\
&&\beta _{2}(\kappa )\equiv 2k\kappa \pm \sqrt{\left[ \kappa ^{2}+2\cos
\delta \right] \left[ \kappa ^{2}+2\cos \delta +2\rho ^{2}(\chi _{1}-\chi )%
\right] }  \label{omega2}
\end{eqnarray}%
%
%
%

We aim to identify parametric domains where the background is subject to the
MI. Due to the Galilean invariance of underlying Eqs. (\ref{eq:NLS_gen1})
and Eq. (\ref{eq:NLS_gen2}), the instability is not affected by boost $k$.
Next, we observe from Eqs. (\ref{omega1}) and (\ref{omega2}) that there are
three different sources of the MI. Firstly, the instability occurs at
\begin{equation}
\chi _{1}+\chi <0  \label{cond1}
\end{equation}%
This is the "standard" (i.e., observed also for the conservative system of %
 {nonlinearly coupled NLSEs, without linear coupling})
instability stemming from Eq. (\ref{omega1}) due to the long-wavelengths
excitations;  {this domain of the parameters is not influenced
by gain/dissipation}.

Another instability domain,%
\begin{equation}
\cos \delta <\max \{0,\rho ^{2}(\chi -\chi _{1})\},  \label{inst2}
\end{equation}%
ensues from Eq. (\ref{omega2}), and
 {linear coupling between NLSEs gives rise to the appearance of this instability domain. Nevertheless, here presence of gain/dissipation ($\delta \ne 0,\pi$) makes the situation significantly
different from that in conservative system ($\delta =0$ or $\delta =\pi$) \cite{lin-coup}, as distinct from the previous case.}
The largest instability growth rate, $\nu =\max_{\kappa }\{$Im$%
 {\beta}(\kappa )\}$, is
\[
\nu \equiv \mathrm{Im}\left\{  {\beta}(\kappa _{m})\right\}
=\rho ^{2}|\chi _{1}-\chi |,~~\kappa _{m}^{2}=\rho ^{2}(\chi -\chi
_{1})-2\cos \delta ,
\]%
in the case
\begin{eqnarray}
 {2\cos \delta <\rho ^{2}(\chi-\chi _{1})},  \label{case1}
\end{eqnarray}%
and
\[
\nu = {2}\sqrt{\left|\left[\cos \delta +\rho^2 (\chi _{1}-\chi
)\right]\cos \delta \right|}, ~~ {\kappa _{m}^{2}=0}
\]%
at
\begin{equation}
 {\rho ^{2}(\chi-\chi_{1} )<2\cos \delta<0}\,\,\,\mathrm{and}%
\,\,\,0<\rho ^{2}(\chi -\chi _{1})< 2\cos \delta<2\rho
^{2}(\chi -\chi _{1}),  \label{case2}
\end{equation}%
cf. Eq. (\ref{case1}). Note that domain (\ref{case2}) disappears in the case
of the equal SPM and XPM nonlinearities, $\chi _{1}=\chi $ (i.e., in the $%
\mathcal{PT}$-symmetric version of the \textit{Manakov's system} \cite%
{Manakov}).

\begin{figure}[b]
\begin{center}
\includegraphics*[width=8.5cm]{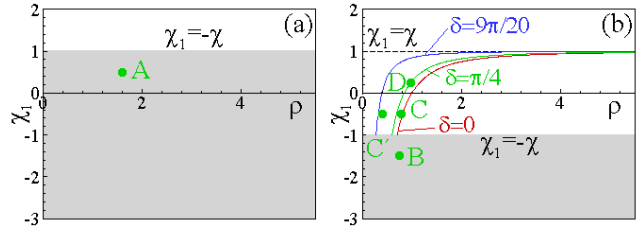}
\end{center}
\caption{Domains of the MI: shaded regions in both panels, as well as the
region under the respective curve, corresponding to different gain/loss
coefficient ($\protect\delta $), in panel (b), in the $(\protect\chi _{1},%
\protect\rho )$-plane for different $\protect\delta $ (as indicated in the
panels) and for fixed XPM coefficients, $\protect\chi =-1$ (a), or $\protect%
\chi =1$ (b). Capital letters indicate parameters chosen for displaying the
evolution in figures following below.}
\label{fig:mi-lim}
\end{figure}

The first consequence of Eq. (\ref{inst2}) is that for $\pi /2\leq \delta
\leq \pi $ (antisymmetric solutions) the background is unstable irrespective
of values of other parameters. The MI regions for symmetric solution ($0\leq
\delta \leq \pi /2$) are displayed in detail Fig.~\ref{fig:mi-lim}, where
the cases of focusing ($\chi =-1$) and defocusing ($\chi =1$) XPM are
considered separately. The former case [Fig.~\ref{fig:mi-lim}(a)] is the
simplest one: here, beyond the fulfillment of condition (\ref{cond1}) [shown
by the shadowed region in Fig.~\ref{fig:mi-lim}(a)], i.e., at $\chi
_{1}>-\chi $, condition (\ref{inst2}) results in $\cos \delta <0$, i.e., it
does not introduce any new domain of the MI. The situation is more
complicated in the case of the defocusing XPM [Fig.~\ref{fig:mi-lim}(b)],
where along, with $\chi _{1}<-\chi $ [the shadowed region], there exists
another MI domain, generated by Eq.~(\ref{inst2}). As a result, the CWs with
large amplitudes, $\rho ^{2}> \cos (\delta) /(2\chi )$, are unstable for $%
\chi _{1}<\chi -\cos (\delta) /\rho ^{2}$. At the same time, at $\delta
\rightarrow \pi /2$, this instability domain approaches the whole area of $%
-\chi <\chi _{1}<\chi $.%

Different origins of the MI should naturally lead to different scenarios of
its development, which we studied by means of direct numerical simulations
of Eqs.~(\ref{eq:NLS_gen1}), (\ref{eq:NLS_gen2}). The simulations were
performed subject to periodic boundary conditions, and with initial
excitation of the CW state (\ref{eq:pl-wave-cur}) by adding random noise
with the amplitude amounting to 1$\%$ of that of the unperturbed background.

\begin{figure}[h]
\begin{center}
\includegraphics{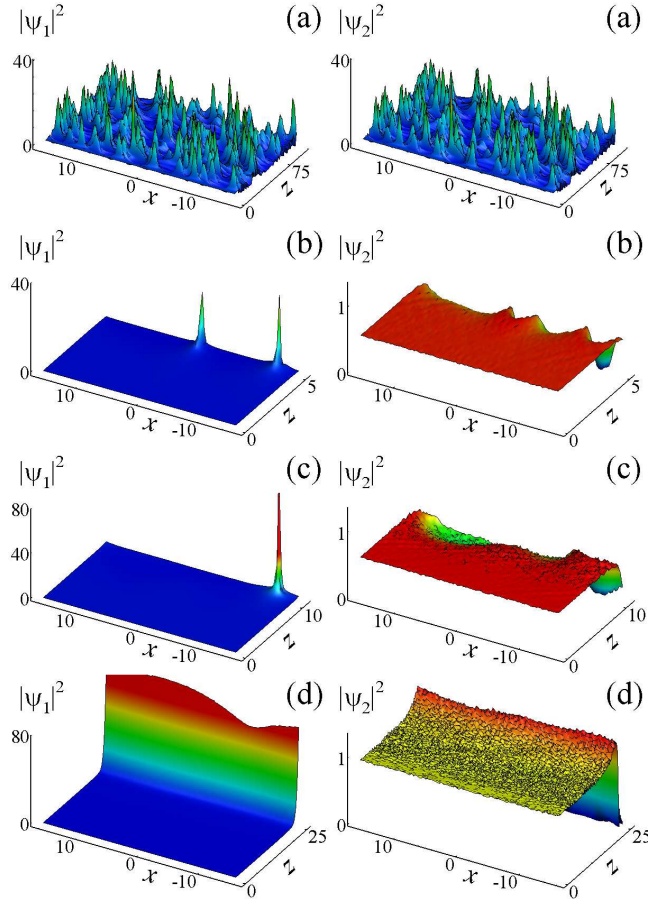}
\end{center}
\caption{The evolution of field components $|\protect\psi_{1}(x,z)|^{2}$ and
$|\protect\psi_{2}(x,z)|^{2}$ (left and right columns) of the plane-wave
solution with parameters $k=0$, $\protect\delta =\protect\pi /4$, $\protect%
\rho =1.604$, $\protect\chi _{1}=0.5$, $\protect\chi =-1$ (a), $\protect\rho %
=0.76$, $\protect\chi _{1}=-1.5$, $\protect\chi =1$ (b), $\protect\rho =0.79$%
, $\protect\chi _{1}=-0.5$, $\protect\chi =1$ (c) and $\protect\rho =0.98$, $%
\protect\chi _{1}=0.25$, $\protect\chi =1$ (d). Parameters of panels (a),
(b), (c) and (d) correspond to points A, B, C and D in Fig. \protect\ref%
{fig:mi-lim}, respectively.}
\label{fig:mi}
\end{figure}

Starting with the case of weak gain and loss, defined by the symmetric
solution ($\delta <\pi /2$), in Fig.~\ref{fig:mi} we show typical results of
these simulations for the focusing XPM [see Fig.~\ref{fig:mi}(a), with
parameters corresponding to point A in Fig.~\ref{fig:mi-lim}(a)], and for
the defocusing XPM [see Figs.~\ref{fig:mi}(b)-(d), with parameters
corresponding to points B-D in Fig.~\ref{fig:mi-lim}(b), respectively]. In
panel (a) we observe a ``standard" scenario of the development of MI. This
behavior, being seemingly expectable, nevertheless reveals a noteworthy
feature of the $\mathcal{PT}$-symmetric system, which behaves similarly to
its Hamiltonian counterpart (at least, in significant initial intervals of
the propagation). In particular, we observe that the power is distributed
between the two waveguides.

The situation changes significantly when one consider the defocusing XPM,
even if Eq. (\ref{cond1}) is satisfied, i.e., the MI has the same nature as
in the conservative system. Indeed, in Fig.~\ref{fig:mi}(b) we observe a
rather fast power transfer from the lossy waveguide to the one with the
gain, accompanied by fast growing peaks. Obviously, such peaks can be
described by a single NLS equation (\ref{eq:NLS_gen1}) with $\psi _{2}=0$.
The observed behavior is due to the focusing SPM, $\chi _{1}$, and therefore
is not significantly altered even when one passes from the domain of
parameters (\ref{cond1}) [Fig.~\ref{fig:mi}(b)] to the one defined by Eq. (%
\ref{inst2}), as shown in Fig.~\ref{fig:mi}(c). A significant change, i.e.,
the third scenario of the evolution of the MI, appears when the SPM is
defocusing too [Fig.~\ref{fig:mi}(d)]. This is the case where the MI occurs
only due to the imbalance of the gain and loss, resulting in nearly
homogeneous grow (decay) of the field in the waveguide with gain
(dissipation), respectively.

\begin{figure}[h]
\begin{center}
\includegraphics{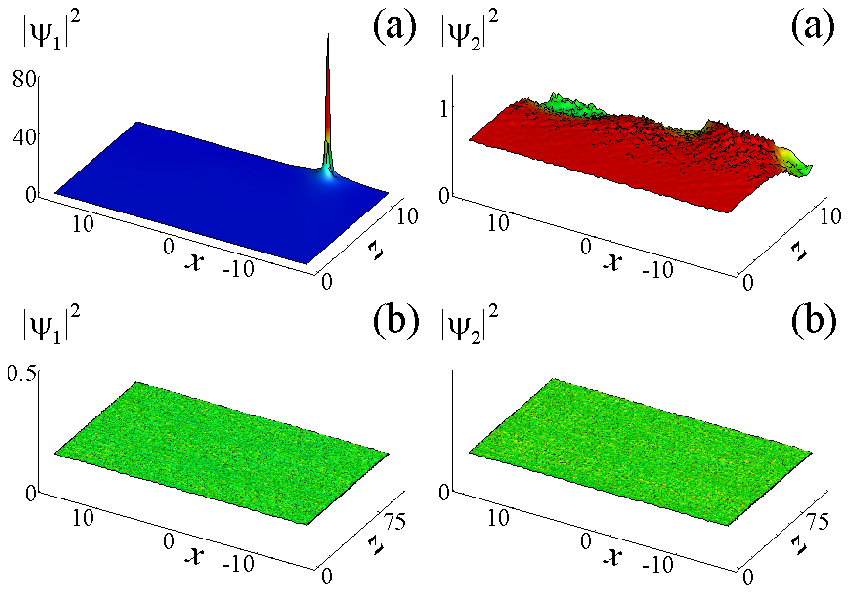}
\end{center}
\caption{The evolution of field components $|\Psi_{1}(x,z)|^{2}$ $%
|\Psi_{2}(x,z)|^{2}$ (left and right columns) of the plane-wave solution
with parameters $k=0.2$, $\protect\delta =\protect\pi /4$, $\protect\chi %
_{1}=-0.5 $, $\protect\chi =1$, $\protect\rho =0.79$ (a), or $\protect\rho %
=0.4$ (b). Parameters of panels (a),(b) correspond to points C, C$^\prime$
in Fig. \protect\ref{fig:mi-lim}(b), respectively. }
\label{fig:mi-k}
\end{figure}

Examples of the modulational instability and stability for the CW solution
with nonzero wave vector $k$ (current) are presented in Fig. \ref{fig:mi-k}.
Here we restrict our consideration of the MI with the focusing SPM, $\chi
_{1}<0$, but when $\chi _{1}+\chi <0$ [point C in Fig.~\ref{fig:mi-lim}(b)].
Thus, the evolution of the MI in this case occurs according to the same
scenario as for $k=0$, cf. Figs .\ref{fig:mi}(c) and Fig.\ref{fig:mi-k}(a).
At the same time, the respective MI peak is shifted in the positive
direction of the $x$-axis, which coincides with the direction of the
current. Meanwhile, in the domain where the CW state is predicted to be
stable [above the green line, in Fig.\ref{fig:mi-lim}(b) --- e.g., at point C%
$^{\prime }$ ], the stability is confirmed by the numerical simulations, see
Fig.\ref{fig:mi-k}(b).

\section{The Peregrine soliton in $\mathcal{PT}$-symmetric system: the case
of $\protect\chi _{1}+\protect\chi <0$}

Turning now  {towards studying} the Peregrine soliton %
 {(rogue wave)} propagating against an unstable background we
start with the case (\ref{cond1}). This readily allows one to write down the
Peregrine solution of Eqs.~(\ref{eq:NLS_gen1}),~(\ref{eq:NLS_gen2}) in the
form ($j=1,2$) \cite{c:rw-BEC2,c:akhmed2011}
\begin{eqnarray}
\psi _{j}(x,z) &=&\rho e^{(-1)^{j}i\delta /2+ikx-ibz}\times   \nonumber \\
&&\left[ 1-4\frac{1-2i\left( \chi _{1}+\chi \right) \rho ^{2}z}{1-2\left(
\chi _{1}+\chi \right) \rho ^{2}(x-2kz)^{2}+4\left( \chi _{1}+\chi \right)
^{2}\rho ^{4}z^{2}}\right] .  \label{eq:peregrine-cur}
\end{eqnarray}%
Notice that, when $|z|\rightarrow \infty $, or, equivalently, $%
|x|\rightarrow \infty $, solution (\ref{eq:peregrine-cur}) merges into the
background given by Eq. (\ref{eq:pl-wave-cur}). Below, we separately
consider two cases: the Peregrine soliton, based on the background without
the current ($k=0$), and current-based Peregrine solution, with $k\neq 0$.
\begin{figure}[tbp]
\includegraphics{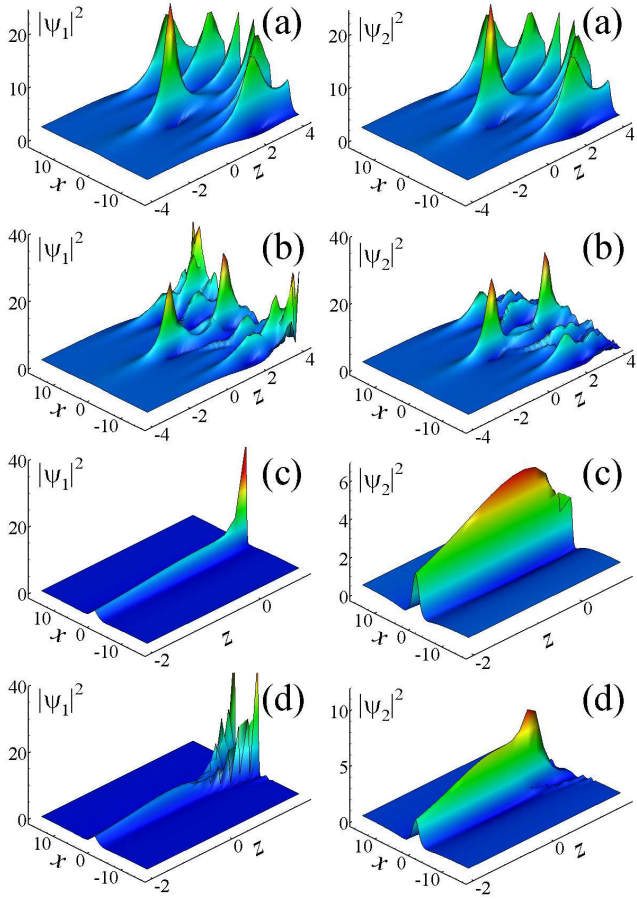}
\caption{Peregrine solutions in the $\mathcal{PT}$-limit for $\protect\rho %
=1.604$, $\protect\chi _{1}=0.5$, $\protect\chi =-1$, $\protect\delta =%
\protect\pi /4$ (a), or $\protect\delta =3\protect\pi /4$ (b); $\protect\rho %
=0.76$, $\protect\chi _{1}=-1.5$, $\protect\chi =1$, $\protect\delta =%
\protect\pi /4$ (c), or $\protect\delta =3\protect\pi /4$ (d). Parameters of
panels (a) and (b) correspond to the point A, while those of panels (c) and
(d) -- to point  {B} in Fig.\protect\ref{fig:mi-lim}.}
\label{fig:rw-hom}
\end{figure}


Examples of the Peregrine solutions whose backgrounds (without the current, $%
k=0$) correspond to points A and  {B} in Fig.~\ref{fig:mi-lim}%
, are depicted in Fig. \ref{fig:rw-hom}. The spatial evolution of $|\psi
_{j}(x,z)|^{2}$ was obtained by the numerical simulations of Eqs.(\ref%
{eq:NLS_gen1}), (\ref{eq:NLS_gen2}) with the initial condition corresponding
to the Peregrine soliton (\ref{eq:peregrine-cur}) at $z=z_{i}=-4$ [Figs. \ref%
{fig:rw-hom}(a) and \ref{fig:rw-hom}(b)], or $z=z_{i}=-2$ [Figs. \ref%
{fig:rw-hom}(c) and \ref{fig:rw-hom}(d)]. In the case of the defocusing SPM
and focusing XPM [Figs.\ref{fig:rw-hom}(a) and \ref{fig:rw-hom}(b)], the
central peak, corresponding to the Peregrine solution, appears at $x=z=0$,
before MI peaks. At the same time, in the case of the focusing SPM and
defocusing XPM [Figs. \ref{fig:rw-hom}(c) and \ref{fig:rw-hom}(d)], the
appearance of the Peregrine-soliton peak at $x=z=0$ causes further growth of
the peak in the first (gain-pumped) component, and decrease in the second
(lossy) core. Notice that the structure of the rogue-wave evolution in this
case resembles the respective scenario of the MI development for the same
parameters, as suggested by the comparison of Figs.~\ref{fig:mi}(b) and
Figs.~\ref{fig:rw-hom}(c).

%

\begin{figure}[tbp]
\includegraphics{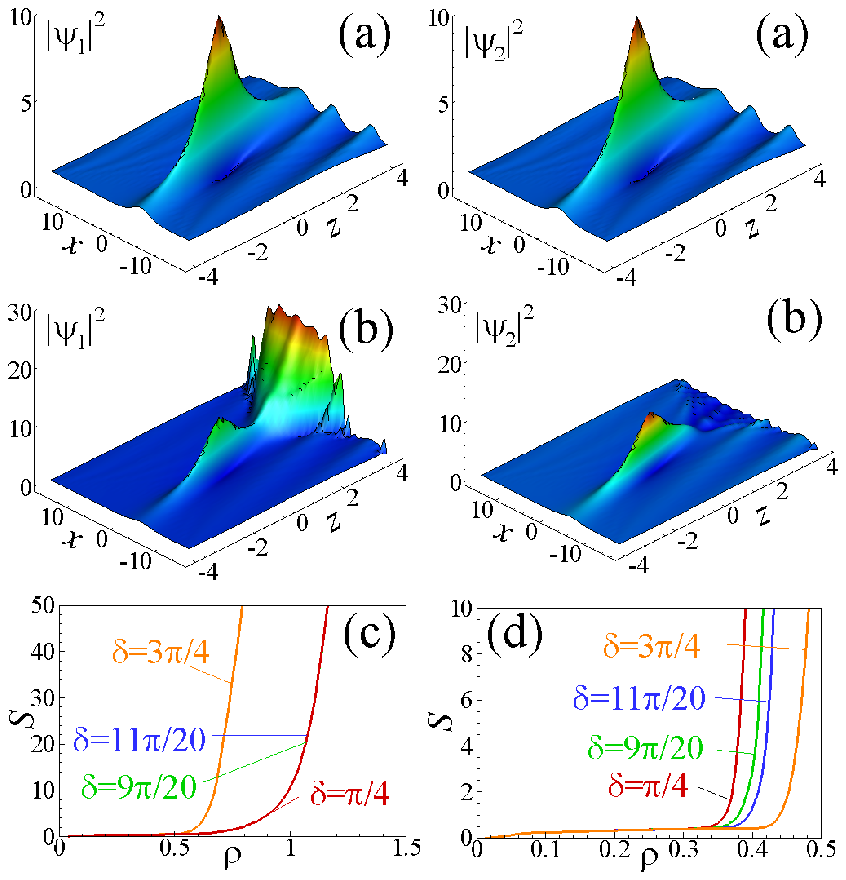}
\caption{(a,b) Current-based Peregrine solutions in the $\mathcal{PT}$%
-symmetric system for $k=0.6$, $\protect\rho =1.0$, $\protect\chi _{1}=0.5$,
$\protect\chi =-1$, and $\protect\delta =\protect\pi /4$ (a), or $\protect%
\delta =3\protect\pi /4$ (b); (c,d) Discrepancy $S$ \textit{vs} amplitude $%
\protect\rho $ for different gain/loss coefficients $\protect\delta $
(indicated in panels) and for $\protect\chi _{1}=0.5$ and the focusing XPM,
with $\protect\chi =-1$ (c) or $\protect\chi _{1}=-1.5$ and the defocusing
XPM, with $\protect\chi =1$.}
\label{fig:rw-k}
\end{figure}

In the case when the background carries the current ($k\neq 0$), the central
peak of the Peregrine solution moves with group velocity $2k$ in the
positive direction of $x$-axis, as seen in Eq.(\ref{eq:peregrine-cur}) and
confirmed by Figs. \ref{fig:rw-k}(a,b). Also for the focusing-XPM ($\chi <0$%
) case, the $\mathcal{PT}$-symmetric ($\delta <\pi /2$) rogue wave is more
``stable" (in the sense that the MI peaks appear after at a longer
propagation distance after the principal rogue-wave peak), see Fig.\ref%
{fig:rw-k}(a), if compared to the $\mathcal{PT}$-antisymmetric wave with $%
\delta >\pi /2$, see Fig. \ref{fig:rw-k}(b).

In order to describe this rogue wave ``stability" quantitatively, we will
use one of the principal properties of Peregrine solution, which follows
from Eq. (\ref{eq:peregrine-cur}), namely $\overline{\psi }_{j}(-x,-z)=\psi
_{3-j}(x,z)$. If the phase of the solution is not taken into account, this
property turns into $|\psi _{j}(-x,-z)|^{2}=|\psi _{j}(x,z)|^{2}$. Thus, we
introduce the discrepancy as
\[
S=\int_{-\infty }^{\infty }\!\left[ \left\vert \psi
_{1}(-x,-z_{i})\right\vert ^{2}+\left\vert \psi _{2}(-x,-z_{i})\right\vert
^{2}-\left\vert \psi _{1}(x,z_{i})\right\vert ^{2}-\left\vert \psi
_{2}(x,z_{i})\right\vert ^{2}\right] ^{2}dx,
\]
in order to eliminate phase effects. In the ideal case, where the shape of
the rogue wave coincides with the Peregrine solution (\ref{eq:peregrine-cur}%
), the discrepancy is zero, $S\equiv 0$. Thus, $S$ serves to measure how
much the numerically obtained solution differs from the Peregrine soliton,
or in other words, how much the  {chaotic nature of} MI %
 {influences} the Peregrine solution. The results are depicted
in Figs. \ref{fig:rw-k}(c) and \ref{fig:rw-k}(d). For the focusing XPM [Fig.%
\ref{fig:rw-k}(c)] and for $\delta \lesssim \pi /2$ the discrepancy abruptly
grows at $\rho \gtrsim 1$. In the same time, in this range of $\delta $ the
discrepancy almost does not depend on $\delta $ [the lines for $\delta =\pi
/4$, $\delta =9\pi /20$, and $\delta =11\pi /20$ are indistinguishable on
the scale of Fig. \ref{fig:rw-k}(c)]. Meanwhile, for $\delta >\pi /2$ the
situation is opposite: the discrepancy increases with $\delta $ (compare the
lines for $\delta =11\pi /20$ and $\delta =3\pi /4$). For the defocusing XPM
[Fig. \ref{fig:rw-k}(d)], discrepancy $S$ decreases with the increase of $%
\delta $ in the whole range of $0\leq \delta \leq \pi $, while an abrupt
growth happens at $\rho \gtrsim 0.4$. As a result, for the focusing XPM, the
$\mathcal{PT}$-symmetric rogue wave with $\delta <\pi /2$ is \emph{more
stable} than the its antisymmetric counterpart, while for the defocusing XPM
the situation is opposite.

\section{Bright solitons}

Obvious bright-soliton solutions of Eq. (\ref{1}) with arbitrary amplitude $%
\eta $ are available too, for $\chi _{1}+\chi <0$:%
\begin{equation}
\psi _{j} =\frac{\eta }{\sqrt{|\chi _{1}+\chi |}\cosh \left( \eta x/\sqrt{2}%
\right) }\exp \left[i\left( (-1)^{j}\frac{\delta}{2}+ \cos \delta +\frac{1}{2%
}\eta ^{2}\right) z\right] ,  \label{sol}
\end{equation}%
$j=1,2$, where, as above, intervals $0\leq \delta \leq \pi /2$ and $\pi
/2\leq \delta \leq \pi $ correspond for the $\mathcal{PT}$-symmetric and
antisymmetric solitons, respectively. 
Using results from Refs. \cite{c:solitons} and \cite{HS}, an \emph{exact}
stability boundary for the symmetric and antisymmetric solitons, against
small perturbations breaking the respective symmetry or antisymmetry, can be
predicted in the following analytical form:%
\begin{equation}
\eta _{\mathrm{cr}}^{2}=\frac{16\left( -\chi _{1}-\chi \right) \cos \delta }{%
\left( \sqrt{-25\chi _{1}+7\chi }-3\sqrt{-\chi _{1}-\chi }\right) \left(
\sqrt{-25\chi _{1}+7\chi }+\sqrt{-\chi _{1}-\chi }\right) },
\label{boundary}
\end{equation}%
the solitons being stable at $\eta ^{2}<\eta _{\mathrm{cr}}^{2}$.

This result makes sense when Eq. (\ref{boundary}) yields a positive value,
otherwise the $\mathcal{PT}$-symmetry-breaking \textit{bifurcation} does not
occur, and the stability may only be studied numerically [in addition to the
instability mode represented by Eq. (\ref{boundary}), other instabilities
are possible too]. In particular, condition (\ref{boundary}) cannot
simultaneously hold for the $\mathcal{PT}$-symmetric and antisymmetric
solitons. Further, because the existence of the solitons of either type
requires $\chi _{1}+\chi <0$, the condition of $\eta _{\mathrm{cr}}^{2}>0$
actually holds for the $\mathcal{PT}$-symmetric solitons at $-\chi
_{1}>-\chi $, and for the $\mathcal{PT}$-antisymmetric solitons --- in the
opposite case, at $-\chi _{1}<-\chi $.

\begin{figure}[tbp]
\includegraphics[width=8.5cm]{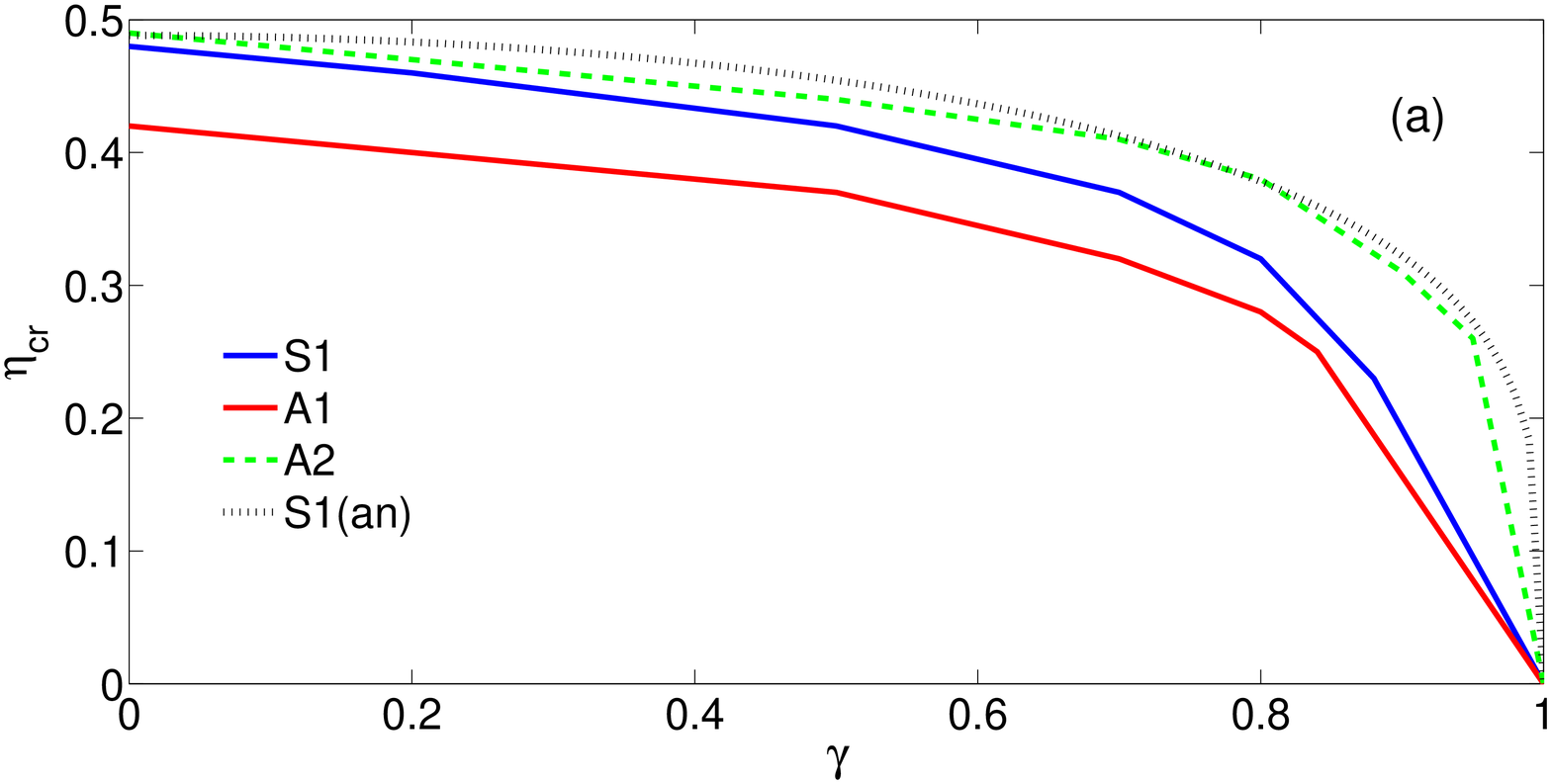} %
\includegraphics[width=8.5cm]{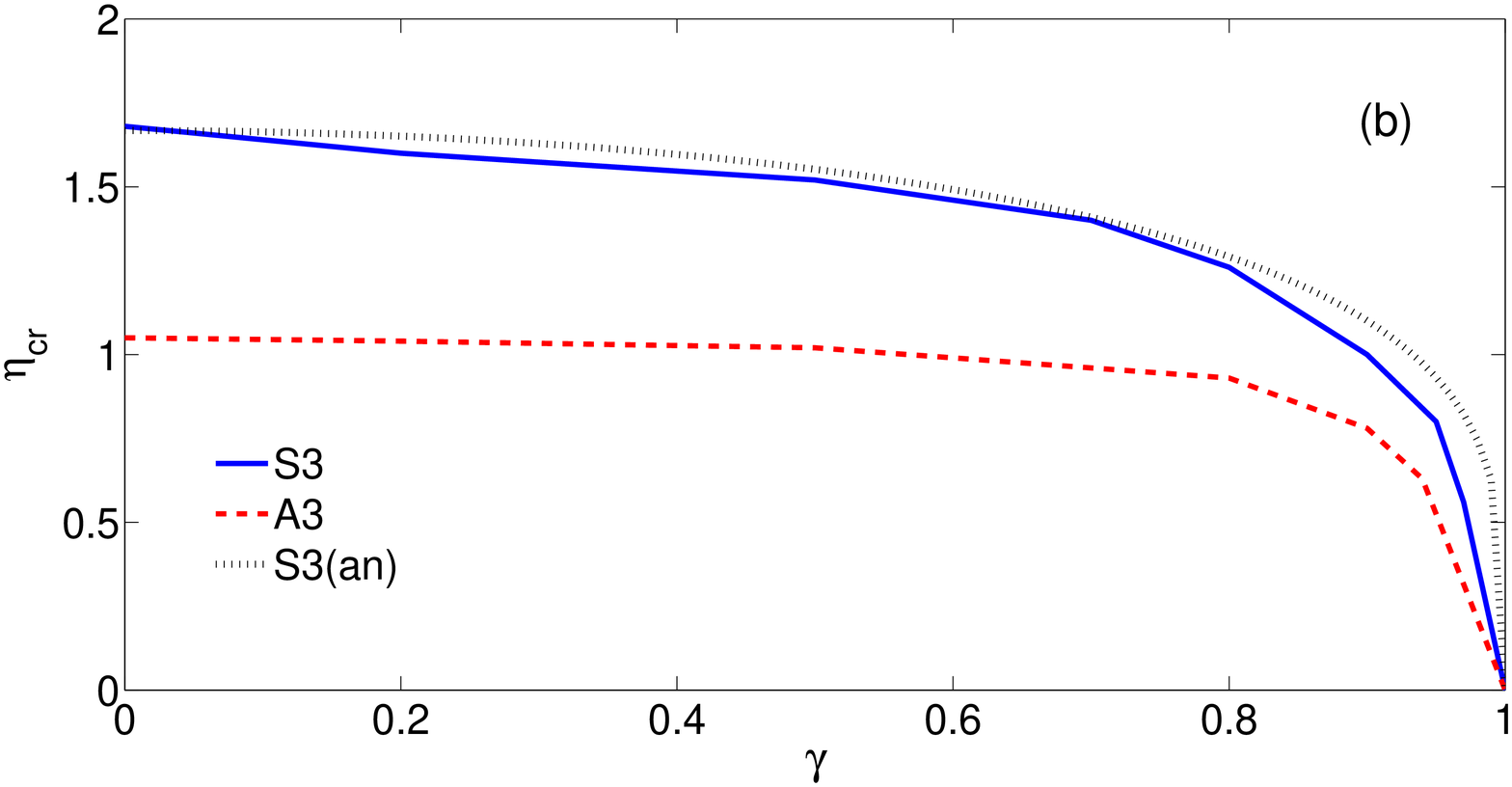}
\caption{(a) Stability boundaries for $\mathcal{PT}$-symmetric and
antisymmetric solitons (\protect\ref{sol}) are shown in the plane of the
gain-loss coefficient $\protect\gamma $ (recall $\protect\gamma \equiv \sin
\protect\delta $) and soliton's amplitude $\protect\eta $, in the case of
opposite signs of the SPM\ and XPM coefficients. The boundaries for the $%
\mathcal{PT}$-symmetric (S1) and antisymmetric (A1) solitons are shown by
solid blue and red lines, respectively, for $\protect\chi =1$ and $\protect%
\chi _{1}=-1.5$. The dotted black curve labeled S1(an) displays the
analytical counterpart of the S1 boundary, as predicted by Eq. (\protect\ref%
{boundary}). The stability boundary for antisymmetric solitons, depicted by
the dashed green curve (A2), pertains to $\protect\chi =-1$ and $\protect%
\chi _{1}=0.5$. (b) The stability boundaries in case of identical signs of
the SPM and XPM coefficients, $\protect\chi =-1$ and $\protect\chi _{1}=-3$.
The boundaries (S3) and (A3) for symmetric and antisymmetric solitons are
shown by solid blue (S) and dashed red curves (A), respectively. The dotted
black curve labeled S3(an) is the analytical counterpart of the latter
boundary, as predicted by Eq. (\protect\ref{boundary}).}
\label{fig:stabilitycurves1}
\end{figure}

We have performed direct simulations of the evolution of perturbed solitons
within the framework of Eqs. (\ref{eq:NLS_gen1}) and (\ref{eq:NLS_gen2}),
aiming to identify stability borders for the $\mathcal{PT}$-symmetric and
anti-symmetric solitons, and, in particular, to verify the analytical
prediction (\ref{boundary}). Perturbations were introduced by adding $2\%$
to the amplitude of component, and reducing $2\%$ from the other. Figures %
\ref{fig:stabilitycurves1}(a) and \ref{fig:stabilitycurves1}(b) display the
so identified stability boundaries in the cases of opposite and identical
signs of $\chi $ and $\chi _{1}$, respectively. For the sake of comparison
with Ref.~\cite{c:solitons,c:PTepl}, we demonstrate these borders as a
function of $\gamma $ , rather than $\delta $ [see Eq. (\ref{delta})].

The numerically found stability boundaries are close to their analytical
counterparts. Some discrepancy between them is explained by the fact that
some solitons, which are stable against infinitesimal perturbations, may be
destabilized by finite-amplitude excitations.

\begin{figure}[tbp]
\includegraphics[width=12cm]{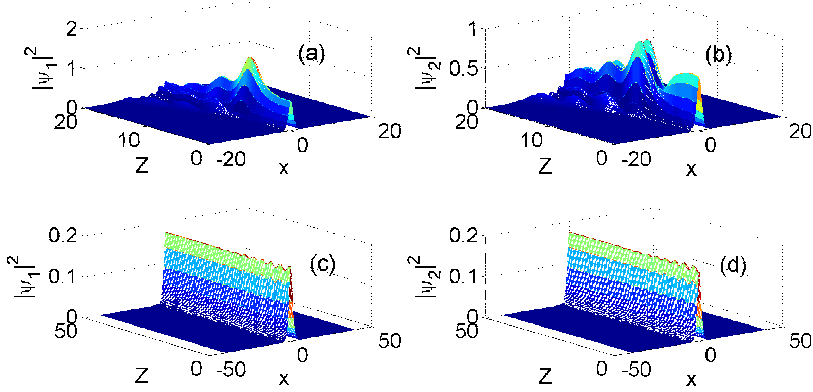}
\caption{Stable and unstable evolution of both components of antisymmetric
solitons (\protect\ref{sol}) at $\protect\chi =-1$ , $\protect\chi _{1}=0.5$
and $\protect\delta =\protect\pi -\arcsin (0.2)$. Panels (a) and (b) pertain
to the unstable dynamics of the soliton with amplitude $\protect\eta =0.6$,
while panels (c) and (d) pertain to the stable soliton with $\protect\eta %
=0.3$.}
\label{fig:examples}
\end{figure}

\begin{figure}[tbp]
\includegraphics[width=8.5cm]{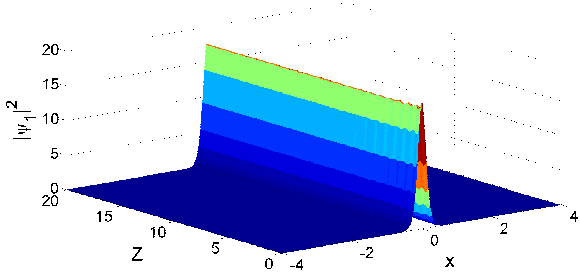}
\caption{Quick stabilization of a symmetric soliton with a large value, $%
\protect\eta =3$, for $\protect\chi =-1$, $\protect\chi _{1}=0.5$ and $%
\protect\gamma =0.2$. Dynamics of the $v$-component is similar to that shown
here for the $u$-component.}
\label{fig:quick}
\end{figure}

Typical examples of the unstable and stable evolution of antisymmetric
solitons, taken on both sides of the stability boundary, are demonstrated in
Fig.\ref{fig:examples}, for $\chi =-1$ and $\chi _{1}=0.5$. The quick
stabilization of the symmetric soliton in the same case is demonstrated in
Fig.\ref{fig:quick} for a large amplitude, $\eta =3$. Actually, the $%
\mathcal{PT}$-symmetric solitons are stable for all $\eta $ in this case,
the stability border being relevant for the antisymmetric ones.

It is relevant to note too that, in the Manakov's limit, $\chi _{1}=\chi $
\cite{Manakov}, the stability boundary predicted by Eq. (\ref{boundary})
diverges. Indeed, direct simulations demonstrate that all the solitons are
stable in this case.

\section{Conclusion}

In this work we have considered the MI (modulational instability) of CW
backgrounds and the emergence and evolution of rogue waves in the system of
linearly-coupled $\mathcal{PT}$-symmetric coupled NLSEs. We have shown that
the focusing XPM nonlinear interactions extend the effective stability
region for the rogue waves of the Peregrine's type. The system can support
nondissipative rogue waves too. The stability region for $\mathcal{PT}$%
-symmetric and antisymmetric solitons was found in the exact analytical form
and verified by direct simulations. It may be interesting to extend the
analysis for  {(2D)} versions of the system, which may have
realizations in nonlinear optics, cf. Ref. \cite{Pavel} and references
therein.

\ack

Y.V.B. acknowledges the support of Funda\c{c}\~ao para a Ci\^encia e a
Tecnologia (Portugal) under Grant No. PEst-C/FIS/UI0607/2011. RD and BAM
appreciate a partial support from grant No. 2010239 provided by the
Binational (US-Israel) Science Foundation. VVK acknowledges support of the
Funda\c{c}\~{a}o para a Ci\^{e}ncia e a Technologia (Portugal) under the
grants PEst-OE/FIS/UI0618/2011 and PTDC/FIS/112624/2009.

\end{document}